\documentclass[twocolumn,showpacs,superscriptaddress,preprintnumbers,amsmath,amssymb,prc]{revtex4-2}

\usepackage{graphicx}
\usepackage{dcolumn}
\usepackage{bm}
\usepackage{float}
\usepackage{color}
\usepackage{ulem}
\usepackage{CJK}
\usepackage[colorlinks,linkcolor=blue,urlcolor=blue,citecolor=blue]{hyperref}

\usepackage{tikz,xcolor,hyperref}

\definecolor{lime}{HTML}{A6CE39}
\DeclareRobustCommand{\orcidicon}{
	\begin{tikzpicture}
	\draw[lime, fill=lime] (0,0) 
	circle [radius=0.16] 
	node[white] {{\fontfamily{qag}\selectfont \tiny ID}};
	\draw[white, fill=white] (-0.0625,0.095) 
	circle [radius=0.007];
	\end{tikzpicture}
	\hspace{-2mm}
}
\foreach \x in {A, ..., Z}{%
	\expandafter\xdef\csname orcid\x\endcsname{\noexpand\href{https://orcid.org/\csname orcidauthor\x\endcsname}{\noexpand\orcidicon}}
}

\begin{document}
\begin{CJK*}{UTF8}{gbsn}

\title{
 Effect of initial-state geometric configurations on the nuclear
   liquid-gas phase transition}
   
\author{Y. T. Cao(曹雅婷)\orcidA{}}
\affiliation{Key Laboratory of Nuclear Physics and Ion-beam Application (MOE), Institute of Modern Physics, Fudan University, Shanghai 200433, China}

\author{X. G. Deng(邓先概)\orcidB{}} 
\email[Corresponding author: ]{ xiangai\_deng@fudan.edu.cn}
\affiliation{Key Laboratory of Nuclear Physics and Ion-beam Application (MOE), Institute of Modern Physics, Fudan University, Shanghai 200433, China}
\affiliation{Shanghai Research Center for Theoretical Nuclear Physics， NSFC and Fudan University, Shanghai 200438, China}

\author{Y. G. Ma(马余刚)\orcidC{}}
\email[Corresponding author: ]{mayugang@fudan.edu.cn}
\affiliation{Key Laboratory of Nuclear Physics and Ion-beam Application (MOE), Institute of Modern Physics, Fudan University, Shanghai 200433, China}
\affiliation{Shanghai Research Center for Theoretical Nuclear Physics， NSFC and Fudan University, Shanghai 200438, China}

\date{\today}

\begin{abstract}

Within the framework of an extended quantum molecular dynamics model, we simulated $^{40}$Ca + $^{16}$O collisions at beam energies ranging from 60 to 150 MeV/nucleon for $^{16}$O with different $\alpha$-cluster configurations. Results imply that different $\alpha$-cluster configurations lead to different yields of deuteron, triton, $^3$He and $^4$He, but not for proton and neutron. We discuss the effect of geometric fluctuations which are presented by double ratios of light nuclei, namely  
$\mathcal{O}_\text{p-d-t}$ and 
$\mathcal{O}_\text{p-d-He}$. 
It is found that magnitude hierarchy of geometric fluctuations is chain, kite, square and tetrahedron structure of $^{16}$O. $\mathcal{O}_\text{p-d-t}$ has maximum value around 80 -- 100 MeV/nucleon which could be  related to liquid-gas phase transition, that is consistent with results from the charge distribution of the heaviest fragments in the collisions.

\end{abstract}


\maketitle

\section{Introduction}
\label{Introduction}

Phase transition is a universal property of interacting substances and generally studied in the thermodynamic limit of macroscopic systems. The atomic nucleus as a finite size system, the phase transition in nucleonic level \cite{Bertsch83,DHEG90,Moretto95,Muller,Bor} or quark level \cite{Gross,de,Aoki,Zhang,Chen,Li} has been extensively discussed and investigated. The interaction between nucleons is similar to that between molecules in a van der Waals fluid, so Bertsch and Siemens~\cite{Bertsch83} speculated that nucleus may experience liquid-gas phase transition (LGPT) when it is heated. Theoretical and experimental efforts were made to confirm it, especially in the area of intermediate energy heavy-ion collisions. In a certain excitation energy range, the nuclear caloric curve has a temperature plateau~\cite{DHEG90}, which implied a possible indication of phase transition~\cite{Moretto95,Natowitz0201,Natowitz0202,YGM97,YGM99,YGM05,YGM20}. Experimentally, spinodal decomposition was found to have occurred in nuclear multifragmentation~\cite{Borderie01}, indicating the existence of liquid-gas phase coexistence region in the finite nuclear systems. The application of negative microcanonical heat capacity in nuclear fragmentation~\cite{Gulminelli02}, which may be related to LGPT~\cite{Chomaz00}.

As we know, clustering is a fundamental phenomenon in physics, which has attracted a lot of attention for a long time. It was  earlier proposed  by Gamow~\cite{Gamow30} and discussed by Bethe and Bacher~\cite{Bethe36,Bethe37} for the high stability of the $\alpha$-cluster around neighboring light nuclei. A cluster structure can emerge in excited states of nuclei or in ground states of nuclei especially in light nuclei, where the nucleus resembles a molecule composed of clusters~\cite{Ikeda,Ortzen,THSR,Freer,Nature,ZhouB,WBH14,HuangBS,WangYZ,LiYA}. Configuration of $\alpha$-cluster is a key problem to understand the phenomenon of clustering in light nuclei. At present, there are many theoretical predictions on $\alpha$-cluster configurations in light nuclei. For instance, $^{16}$O can be treated as linear-chain structure with four-$\alpha$ clusters, which was supported by the $\alpha$ cluster model~\cite{Bauhoff35} and the cranked Skyrme Hartree-Fock method~\cite{Ichikawa11}. At the ground state, it can be regarded as tetrahedral structure with the approach of nuclear chiral effective field theory~\cite{Epelbaum14} and covariant density functional theory~\cite{Liu12}. And the same structure is also presented above the ground state supported by Hartree-Fock-Bogoliubov method~\cite{Girod13}.

In the last decade, many studies have focused on density fluctuations to investigate LGPT as in Refs.~\cite{Steinheimer12,Steinheimer13,Steinheimer16}. Obviously, different $\alpha$-cluster configurations shall induce different geometric fluctuations, so we chose the following four $\alpha$-cluster configurations for the projectile $^{16}$O, which are chain, kite, square and tetrahedron to probe density fluctuation.  How different $\alpha$-cluster configurations affect on the LGPT is considered in this work. In our study, we explore the effect of geometric fluctuation on LGPT in low-intermediate energy heavy-ion collisions. Within the framework of the extended quantum molecular dynamics (EQMD) model, the central $^{40}$Ca + $^{16}$O collisions at energies ranging from 60 to 150 MeV/nucleon are simulated, and the GEMINI model~\cite{GMY19,HGC21,LML22} is then used to de-excite heavy fragments.

The organization of the paper is as follows: In Sect.~\ref{Model&methodology}, we give introductions of our simulation model and method, including the EQMD model and GEMINI model as well as ratios of light nuclei. Results of effects of geometric fluctuation on  the yields and (double) ratios of light nuclei  are discussed in Sect.~\ref{Results&Discussion}. Moreover, the relation to nuclear liquid gas phase transition is pointed out by the charge distribution of the heaviest fragments in the same collisions. Finally, conclusion is given in Sect.~\ref{Conclusion}.

\section{Model and methodology}
\label{Model&methodology}

\subsection{EQMD model}
\label{EQMD model}

In the EQMD model, the wave packets of nucleons are Gaussian-like and the total wave function of the system is treated as the direct product of all nucleons~\cite{Maruyama96}
\begin{equation}
\Psi = \prod_{i} \varphi(\boldsymbol{r}_i)     
=\prod_{i} (\frac{\nu_{i}+\nu^{\ast}_{i}}{2\pi})^{3/4}{\rm exp}[-\frac{\nu_{i}}{2}(\boldsymbol{r}_i-\boldsymbol{R}_i)^2+\frac{i}{\hbar}\boldsymbol{P}_i\cdot\boldsymbol{r}_i]     \,       ,
\label{Wavepacket02}
\end{equation}
where $\boldsymbol{R}_i$ and $\boldsymbol{P}_i$ are the centers of position and momentum of the $i$-th wave packet, respectively. The Gaussian width $\nu_{i}$ is introduced as
$\nu_{i}\equiv\frac{1}{\lambda_{i}}+i\delta_{i}$
where $\lambda_{i}$ and $\delta_{i}$ are dynamical variables in the process of initialization.

The expected value of Hamiltonian can be expressed as
\begin{equation}
\begin{split}
H& = \left\langle\Psi\left|\sum_{i}-\frac{\hbar^{2}}{2m}\bigtriangledown^{2}_{i}-\hat{T}_{zero}+\hat{H}_{int}\right|\Psi\right\rangle\\
& = \sum_{i}\frac{\boldsymbol{P}^{2}_{i}}{2m}+\frac{3\hbar^{2}(1+\lambda^{2}_{i}\delta^{2}_{i})}{4m\lambda_{i}}-T_{zero}+H_{int},
\label{Hamiltonian}
\end{split}
\end{equation}
where the first, second and third term are the center momentum of the wave packet, the contribution of the dynamic wave packet, and the zero point center-of-mass kinetic energy $-T_{zero}$, respectively. The first term can be expressed as $\left\langle\hat{\boldsymbol{p}}_{i}\right\rangle^{2}/2m$,  the second term can be treated as $\left(\left\langle\hat{\boldsymbol{p}}_{i}^{2}\right\rangle-\left\langle\hat{\boldsymbol{p}}_{i}\right\rangle^{2}\right)/2m$, and the form of the third term can be found in details in Ref.~\cite{Maruyama96}.

For the effective interaction $H_{int}$, it consists of the Skyrme potential, the Coulomb potential, the symmetry energy, and the Pauli potential as follows
\begin{equation}
H_{int} = H_{Skyrme} + H_{Coulomb} + H_{Symmetry} + H_{Pauli}.
\label{Effectiveinteraction}
\end{equation}

The form of Skyrme interaction is written as
\begin{equation}
H_{Skyrme} = \frac{\alpha}{2\rho_{0}}\int\rho^{2}(\boldsymbol{r})d^{3}r+\frac{\beta}{(\gamma+1)\rho_{0}^{\gamma}}\int\rho^{\gamma+1}(\boldsymbol{r})d^{3}r,
\label{Skyrmeinteraction}
\end{equation}
where $\alpha = -124.3$ MeV, $\beta = 70.5$ MeV, and $\gamma = 2$, which can be obtained from fitting the ground state properties of finite nuclei.

The form of Coulomb potential can be expressed as
\begin{equation}
H_{Coulomb} = \frac{e^{2}}{2}\sum_{i}\sum_{i\neq{j}}Z_{i}Z_{j}\frac{1}{r_{ij}}{\rm erf}(\frac{r_{ij}}{\sqrt{4L}})     \,      ,
\label{EQMDCoulombpotential01}
\end{equation}
where
$r_{ij} = |\boldsymbol{r}_{i}-\boldsymbol{r}_{j}| $
and 
${\rm erf}(x) = \frac{2}{\sqrt{\pi}}\int^{x}_{0}e^{-u^{2}}du     $.

And the symmetry potential can be written as
\begin{equation}
H_{Symmetry} = \frac{C_{S}}{2\rho_{0}}\sum_{i,j\neq{i}}\int[2\delta(I_{i},I_{j})-1]\rho_{i}(\boldsymbol{r})\rho_{j}(\boldsymbol{r})d^{3}r,
\label{Symmetrypotential}
\end{equation}
where $C_{S}$ is the symmetry energy coefficient which is 25 MeV in this work.

It is known that the stability of nuclei in the model description is very important to study the cluster structure effects of nuclei. As a result, in order to make saturation property and $\alpha$-cluster structures can be obtained after energy cooling~\cite{WBH14}, a phenomenological repulsive Pauli potential is introduced to prevent nucleons with the same spin-$S$ and isospin-$I$ to come close to each other in the phase space, which can be presented as
\begin{equation}
H_{Pauli} = \frac{c_{P}}{2}\sum_{i}(f_{i}-f_{0})^{\mu}\theta(f_{i}-f_{0}),
\label{Paulipotential01}
\end{equation}
where $f_{i}$ is 
the overlap of the $i$-th nucleon with other nucleons having the same spin and isospin, i.e. 
$f_{i}\equiv\sum_{j}\delta(S_{i},S_{j})\delta(I_{i},I_{j})\left|\left\langle\phi_{i}|\phi_{j}\right\rangle\right|^{2}$,
and 
$\theta$ is the unit step function, and $c_{P} = 15$ MeV is a coefficient denoting strength of Pauli potential. For the other two parameters, we take $f_{0} = 1.0$ and $\mu = 1.3$.

For the standard QMD model, it shows insufficient stability, for which the phase space obtained from the Monte Carlo samples is not in the lowest point of energy~\cite{Maruyama96}. So the EQMD model takes the kinetic-energy term of the momentum variance of wave packets in the Hamiltonian into account, which is ignored as the spurious constant term in the standard QMD~\cite{Aichelin86,Hartnack98}. Besides, the wave packet width is introduced into the Hamiltonian as a complex variable, and treated as an independent dynamic variable. These modifications not only describe the ground state better, but also make the model successful in the study of nuclear cluster states.

As a consequence, we first consider that the energy-minimum state is the ground state of initial nucleus. Afterwards, a random configuration is given to each nucleus. And under the time-dependent variation principle (TDVP)~\cite{Kerman76}, propagation of each nucleon can be described as~\cite{Maruyama96}
\begin{equation}
\begin{split}
\boldsymbol{\dot{R}}_{i} = \frac{\partial{H}}{\partial{\boldsymbol{P}_{i}}}+\mu_{\boldsymbol{R}}\frac{\partial{H}}{\partial{\boldsymbol{R}_{i}}},\boldsymbol{\dot{P}}_{i} = -\frac{\partial{H}}{\partial{\boldsymbol{R}_{i}}}+\mu_{\boldsymbol{P}}\frac{\partial{H}}{\partial{\boldsymbol{P}_{i}}},\\
\frac{3\hbar}{4}\dot{\lambda}_{i} = -\frac{\partial{H}}{\partial{\delta_{i}}}+\mu_{\lambda}\frac{\partial{H}}{\partial{\lambda_{i}}},\frac{3\hbar}{4}\dot{\delta}_{i} = \frac{\partial{H}}{\partial{\lambda_{i}}}+\mu_{\delta}\frac{\partial{H}}{\partial{\delta_{i}}},
\label{Dampedequations}
\end{split}
\end{equation}
where $H$ is the expected value of the Hamiltonian, and $\mu_{\boldsymbol{R}}$, $\mu_{\boldsymbol{P}}$, $\mu_{\lambda}$ and $\mu_{\delta}$ are various friction coefficients. During the friction cooling process, the system dissipates its energy with negative coefficients, making itself goes to a stable (minimum or even eigenstate) state~\cite{SSW17}. In contrast, in the subsequent nuclear reaction simulation stage, these coefficients are zero to maintain the energy conservation of the system. ~It is worth mentioning that an improvement in the performance of the inelastic process, especially for the incoherent $p$-$n$ bremsstrahlung process in the framework of the EQMD model, has been presented in Refs.~\cite{CZS20,CZS21}.

\subsection{GEMINI model}
\label{GEMINI model}

The calculation in this study is a two-step process, including both dynamical and statistical codes. At the end of dynamical evolution, the nucleons are re-aggregated and condensed to form individual clusters~\cite{HGC21}. The deexcitation of heavy clusters is realized by the GEMINI code by R. J. Charity~\cite{RJ88,RJ10}. With the information of a given primary fragment including its proton number $Z$, mass number $A$, excitation energy $E^{*}$, and spin $J_{CN}$, GEMINI de-excites the fragment through a series of sequential binary decays until the excitation energy of the hot fragments reaches zero. The GEMINI model deals with the evaporation of light particles in the Hauser-Feshbach form~\cite{Hauser52}. The partial decay width of a compound nucleus for the evaporation of particle $i$ is expressed as
\begin{equation}
\begin{split}
\Gamma_{i}(E^{*},J_{CN}) = \frac{1}{2\pi\rho_{CN}(E^{*},J_{CN})}\int d\varepsilon \sum_{J_{d}=0}^{\infty} \sum_{J=\lvert J_{CN}-J_{d} \rvert}^{J_{CN}+J_{d}} \\ \times \sum_{\ell=\lvert J-S_{i} \rvert}^{J+S_{i}} T_{\ell}(\varepsilon)\rho_{d}(E^*-B_i-\varepsilon,J_d)      \,      ,
\label{Gemini01}
\end{split}
\end{equation}
where $J_d$, $S_i$, $J$, and $\ell$ are spin of the daughter nucleus, the spin, the total angular momentum, and the orbital angular momenta of the evaporated particle, respectively; $\varepsilon$ and $B_i$ are respectively its kinetic and separation energy; $T_\ell$ is its transmission coefficient or barrier penetration factor, and $\rho_d$ and $\rho_{CN}$ are respectively the level density of the daughter and compound nucleus.

The description of intermediate-mass fragment emission follows the Moretto form~\cite{Moretto75,Moretto88}, which has been further extended to the following form
\begin{equation}
\begin{split}
\Gamma_{Z,A} = \frac{1}{2\pi\rho_{CN}(E^{*},J_{CN})}\times d\varepsilon \rho_{sad} \\ (E^*-B_{Z,A}(J_{CN})-\varepsilon,J_{CN})      \,      ,
\label{Gemini02}
\end{split}
\end{equation}
where $\rho_{sad}$ is the level density at the saddle point, $\varepsilon$ is the kinetic energy in the fission degree of freedom at the saddle point, $B_{Z,A}(J_{CN})$ is the conditional barrier depending on both the mass and charge asymmetries, and can be expressed as
\begin{equation}
B_{Z,A}(J_{CN}) = B_A ^{Sierk} (J_{CN})+\Delta M+\Delta E_{Coul}-\delta W-\delta P     \,      ,
\label{Gemini03}
\end{equation}
where $\Delta M$ and $\Delta E_{Coul}$ are the mass and Coulomb corrections accounting for the different $Z$ and $A$ values of the two fragments, $\delta W$ and $\delta P$ are the ground-state shell and pairing
corrections to the liquid drop barrier. The quantity $B_A ^{Sierk}$ is the interpolated Sierk barrier for the specified mass asymmetry.

For the symmetric divisions in heavy nuclei, the GEMINI model uses the Bohr-Wheeler form~\cite{Bohr39} to predict the total symmetric fission yield
\begin{equation}
\begin{split}
\Gamma_{BW} = \frac{1}{2\pi\rho_{CN}(E^{*},J_{CN})}\times d\varepsilon \rho_{sad} \\ (E^*-B_{f}(J_{CN})-\varepsilon,J_{CN})      \,      ,
\label{Gemini04}
\end{split}
\end{equation}
where $B_{f}(J_{CN})$ is the spin-dependent fission barrier, read as
\begin{equation}
B_f(J_{CN}) = B_f ^{Sierk} (J_{CN})-\delta W-\delta P     \,      .
\label{Gemini05}
\end{equation}

\subsection{Ratios and density fluctuation}
\label{Ratios and density fluctuation}

In the analytical coalescence formula COAL-SH~\cite{KJS1701} for cluster production, the yield $N_c$ of a cluster at midrapidity and consisting of $A$ constituent particles from the hadronic matter at kinetic freeze-out or emission source of effective temperature $T_{eff}$, volume $V$, and number $N$ of the $i$-th constituent with mass $m_i$ can be read as
\begin{equation}
\begin{split}
N_c = g_{rel}g_{size}g_c M^{3/2} \left[ \prod^{A} _{i=1} \frac{N_i}{m_{i}^{3/2}} \right] \\ \times \prod^{A-1} _{i=1} \frac{(4\pi/\omega)^{3/2}}{V\chi(1+\chi^2)} \left( \frac{\chi^2}{1+\chi^2} \right)^{l_i} G(l_i,\chi)     \,      .
\label{Ratios01}
\end{split}
\end{equation}

In Eq.~(\ref{Ratios01}), $M = \Sigma^A _{i=1} m_i$ is the rest mass of the cluster, $l_i$ is the orbital angular momentum associated with the $i$-th relative coordinate, $\omega$ is the oscillator frequency of the cluster’s internal wave function and is inversely proportional to $Mr^2 _{rms}$ with $r_{rms}$ being the root-mean-square (RMS) radius of the cluster, and $G(l,x) = \Sigma^{l}_{k=0}\frac{l!}{k!(l-k)!}\frac{1}{(2k+1)\chi^{2k}}$ with $\chi = (2T_{eff}/\omega)^{1/2}$ is the suppression factor due to the orbital angular momentum on the coalescence probability~\cite{SC1101,SC1102}. Additionally, $g_c = (2S+1)/(\Pi^{A}_{i=1}(2s_i+1))$ is the coalescence factor for constituents of spin $s_i$ to form a cluster of spin $S$, $g_{rel}$ is the relativistic correction to the effective volume in momentum space, and $g_{size}$ is the correction due to the finite size of produced cluster.

Taking density fluctuations of nucleons into account, the neutron and proton densities in the emission source can be expressed as~\cite{KJS1702,XGD20}
\begin{equation}
n(\vec{r}) = \frac{1}{V} \int n(\vec{r})d\vec{r}+\delta n(\vec{r}) = \left \langle n \right \rangle +\delta n(\vec{r})    \,      .
\label{Ratios04}
\end{equation}
\begin{equation}
n_p (\vec{r}) = \frac{1}{V} \int n_p(\vec{r})d\vec{r}+\delta n_p(\vec{r}) = \left \langle n_p \right \rangle +\delta n_p(\vec{r})     \,      ,
\label{Ratios05}
\end{equation}
where $\langle \cdots \rangle$ represents the average value over space and $\delta n(\vec{r})$($\delta n_p(\vec{r})$) with $\langle \delta n(\vec{r}) \rangle=0$($\langle \delta n_p (\vec{r}) \rangle=0$) represents the fluctuation of neutron (proton) density from its average value $\left \langle n \right \rangle$($\left \langle n_p \right \rangle$). 
Then yields of deuterons and tritrons can be 
approximately written in an analytical coalescence framework as ~\cite{KJS1702} 
\begin{equation}
\begin{split}
N_d = \frac{3}{2^{1/2}} (\frac{2\pi}{m_0 T_{eff}})^{3/2} \int d\vec{r} n(\vec{r}) n_p(\vec{r}) \\
= \frac{3}{2^{1/2}} \left( \frac{2\pi}{m_0 T_{eff}} \right)^{3/2} N_p \langle n \rangle (1+\alpha \Delta n)     \,      ,
\label{Ratios06}
\end{split}
\end{equation}
\begin{equation}
\begin{split}
N_t = \frac{3^{3/2}}{4} (\frac{2\pi}{m_0 T_{eff}})^{3} \int d\vec{r} n(\vec{r})^2 n_p(\vec{r}) \\
= \frac{3^{3/2}}{4} \left( \frac{2\pi}{m_0 T_{eff}} \right)^{3} N_p \langle n \rangle^2 [1+(1+2\alpha)\Delta n]     \,      ,
\label{Ratios07}
\end{split}
\end{equation}
where 
$\alpha$ being the correlation coefficient. In addition, $\Delta n = \langle (\delta n)^2 \rangle / \langle n \rangle^2$ is a dimensionless quantity that characterizes the relative density fluctuation of neutrons.

Combining Eq.~(\ref{Ratios06}) and Eq.~(\ref{Ratios07}), an important double  ratio can be defined as~\cite{KJS1702,XGD20}
\begin{equation}
O_1 \equiv
\mathcal{O}_\text{p-d-t}= \frac{N_p N_t}{N_{d}^{2}}=g \frac{1+(1+2\alpha)\Delta n}{1+\alpha \Delta n}^2     \,      ,
\label{Ratios08}
\end{equation}
with $g = 4/9 \times (3/4)^{1.5} \approx 0.29$. When $\alpha \Delta n$ is much smaller than unity, the correction from $\alpha$ in Eq.~(\ref{Ratios08}) is second-order~\cite{KJS1702}, and $O_1$ can be approximated as
\begin{equation}
O_1 \approx 0.29(1+\Delta n)     \,      .
\label{Ratios09}
\end{equation}
In this way, $O_1$ has a very simple linear dependence on $\Delta n$. We can suggest that the yield ratio of light nuclei can be taken as a direct probe of the large density fluctuations which might be associated with critical phenomenon~\cite{KJS1702}.

Besides, 
another  double ratio of light-nuclei which $\alpha$-particle is involved was also proposed
~\cite{Shuryak20} as
\begin{equation}
O_4 \equiv \mathcal{O}_\text{p-d-He} = \frac{N_{^{4}He} N_{p}^{2}}{N_{d}^{3}}     \,      .
\label{Ratios12}
\end{equation}
From the results in Ref.~\cite{Shuryak20}, it is thought that the above ratio could be taken as a potential probe of critical phenomenon \cite{Sun,STAR_lightnuclei,Ko,Zheng}. From the statistical point of view, the ratios of O$_{1}$ and O$_{4}$ can be considered in this work. Moreover, in our simulations, some single ratios such as $N_n / N_p$ and $N_{^{4}He} / N_{^{3}He}$ are also considered.


\section{Results and discussion}
\label{Results&Discussion}

\begin{figure}[htb]
\setlength{\abovecaptionskip}{0pt}
\setlength{\belowcaptionskip}{0pt}
\centering\includegraphics[scale=0.4]{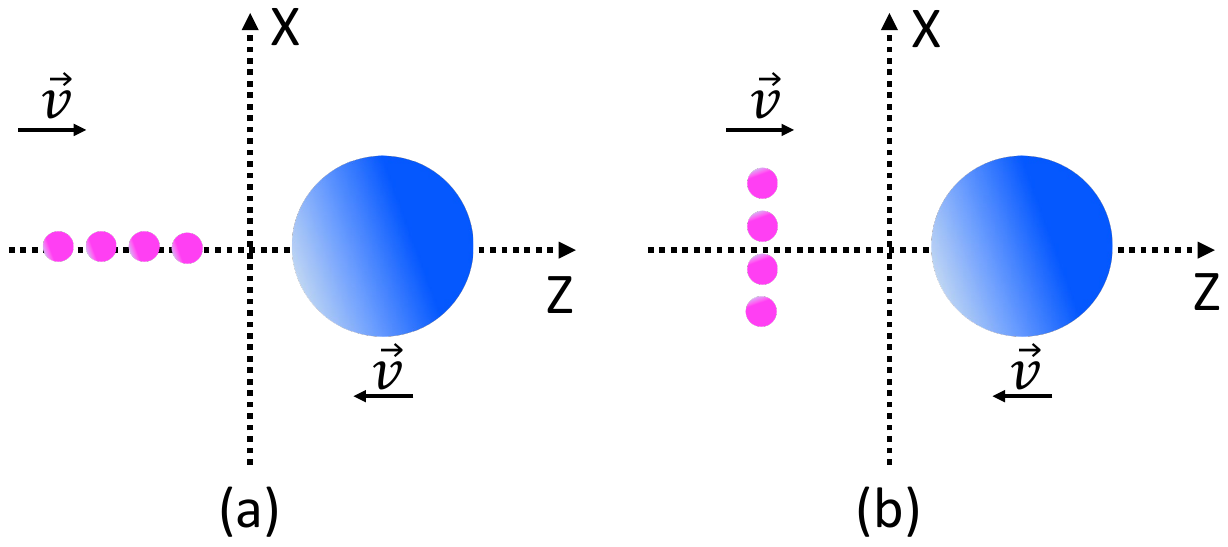}
\caption{Schematic plot of the projections of (a) a transversely polarized and (b) a longitudinally polarized chain-like $^{16}$O in the $x$-$z$ plane at initial stage.}
\label{fig1:transverse}
\end{figure}

\begin{figure*}[htb]
\setlength{\abovecaptionskip}{0pt}
\setlength{\belowcaptionskip}{0pt}
\centering\includegraphics[scale=1.05]{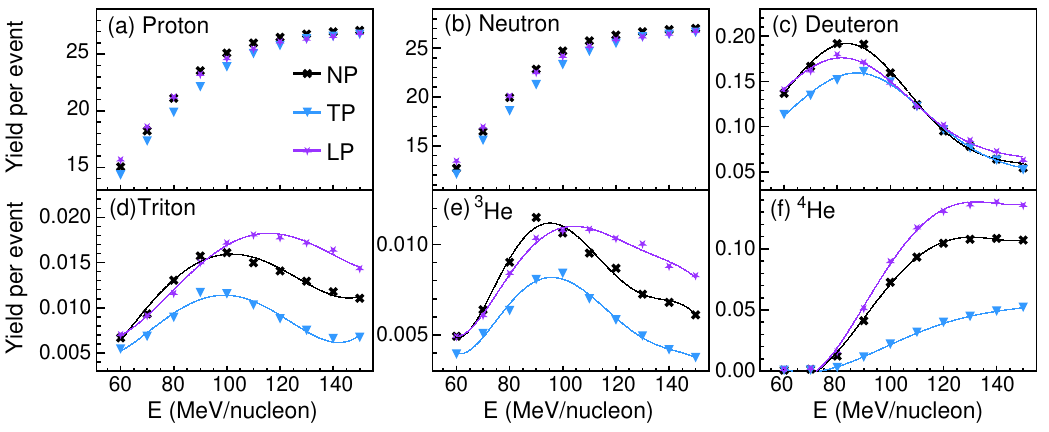}
\caption{Dependence of yields of (a) proton, (b) neutron, (c) deuteron, (d) triton, (e) $^3$He and (f) $^4$He on the incident energy when $^{16}$O is polarized transversely (TP), longitudinally (LP) and unpolarized (NP), respectively.}
\label{fig3:yield_test}
\end{figure*}

\begin{figure}[htb]
\setlength{\abovecaptionskip}{0pt}
\setlength{\belowcaptionskip}{0pt}
\centering\includegraphics[scale=0.55]{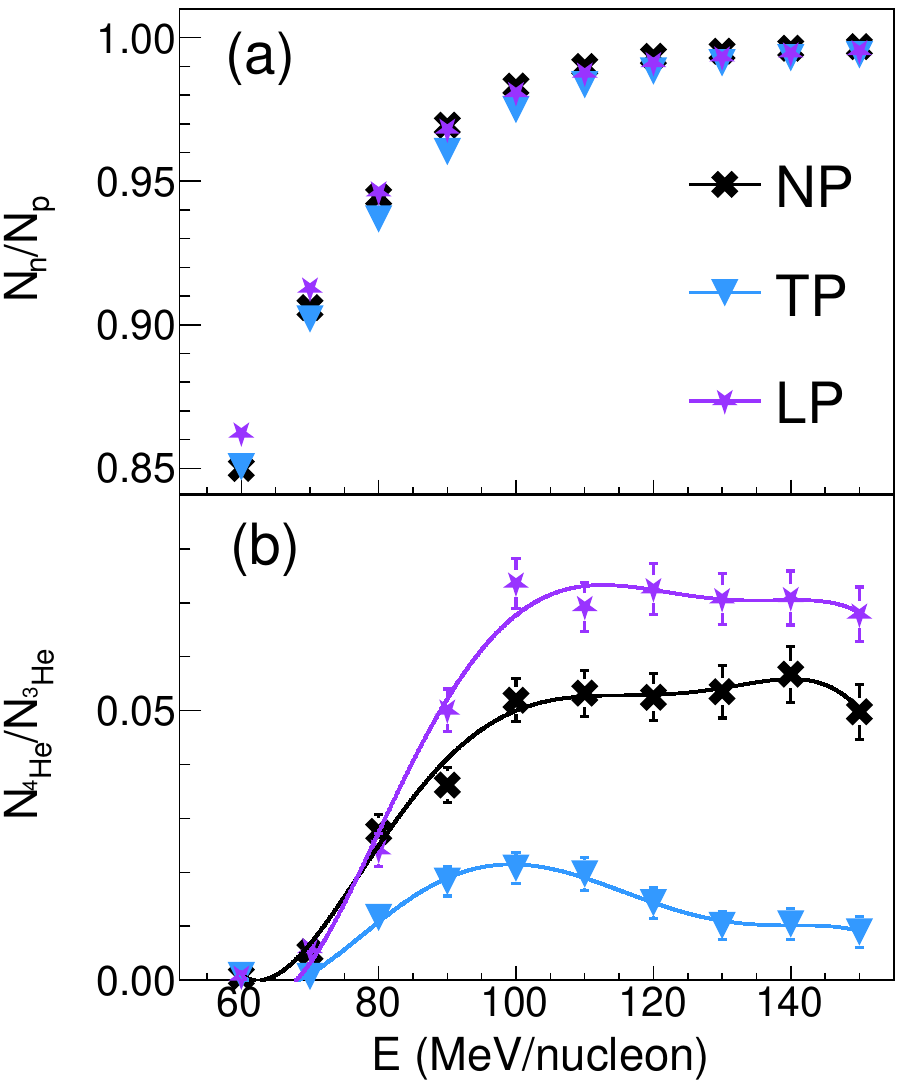}
\caption{Dependence of (a) N$_n$/N$_p$ and (b) N$_{^4He}$/N$_{^3He}$ on the incident energy when $^{16}$O is polarized transversely (TP), longitudinally (LP) and unpolarized (NP), respectively.}
\label{fig5:R_test}
\end{figure}

\begin{figure}[htb]
\setlength{\abovecaptionskip}{0pt}
\setlength{\belowcaptionskip}{0pt}
\centering\includegraphics[scale=0.56]{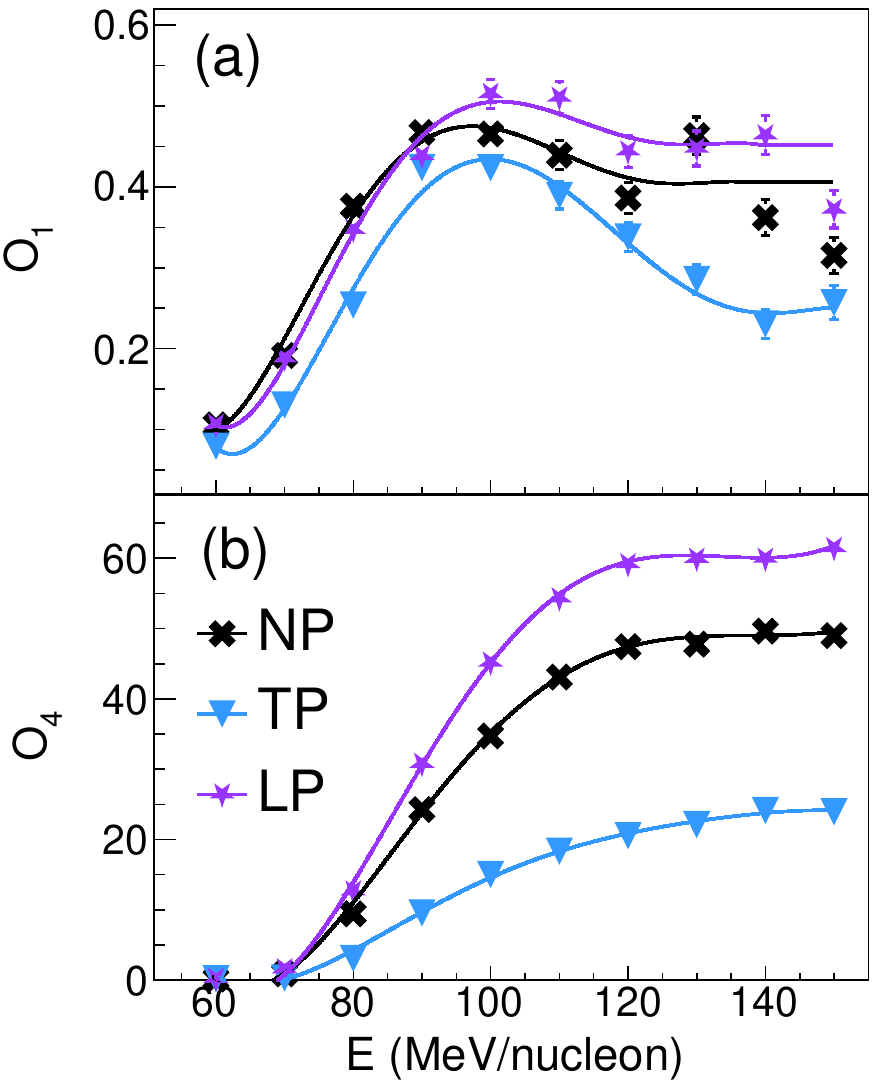}
\caption{Dependence of (a) O$_1$ and (b) O$_4$ on the incident energy when $^{16}$O is polarized transversely (TP), longitudinally (LP) and unpolarized (NP), respectively.}
\label{fig4:O_test}
\end{figure}

\begin{figure*}[htb]
\setlength{\abovecaptionskip}{0pt}
\setlength{\belowcaptionskip}{0pt}
\centering\includegraphics[scale=1.05]{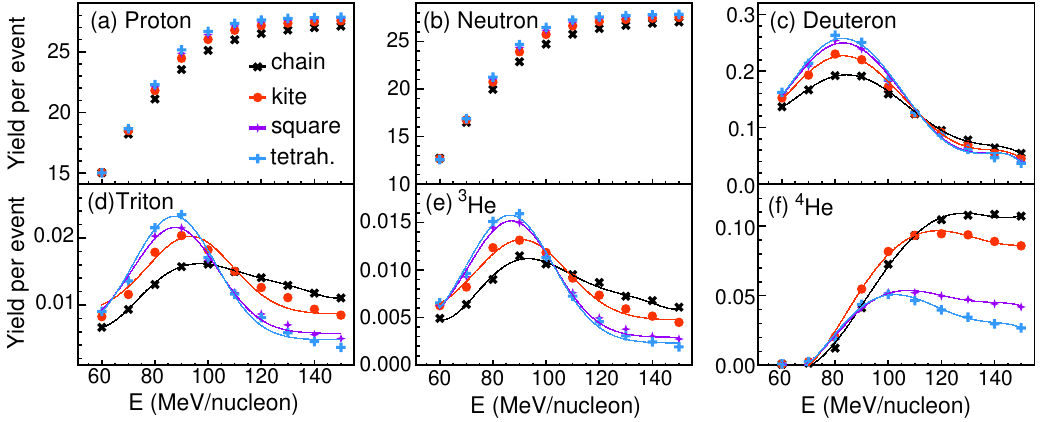}
\caption{Dependence of yields of (a) proton, (b) neutron, (c) deuteron, (d) triton, (e) $^3$He and (f) $^4$He on the incident energy when $^{16}$O has four different $\alpha$-cluster configurations.}
\label{fig6:yield}
\end{figure*}

\begin{figure}[htb]
\setlength{\abovecaptionskip}{0pt}
\setlength{\belowcaptionskip}{0pt}
\centering\includegraphics[scale=0.56]{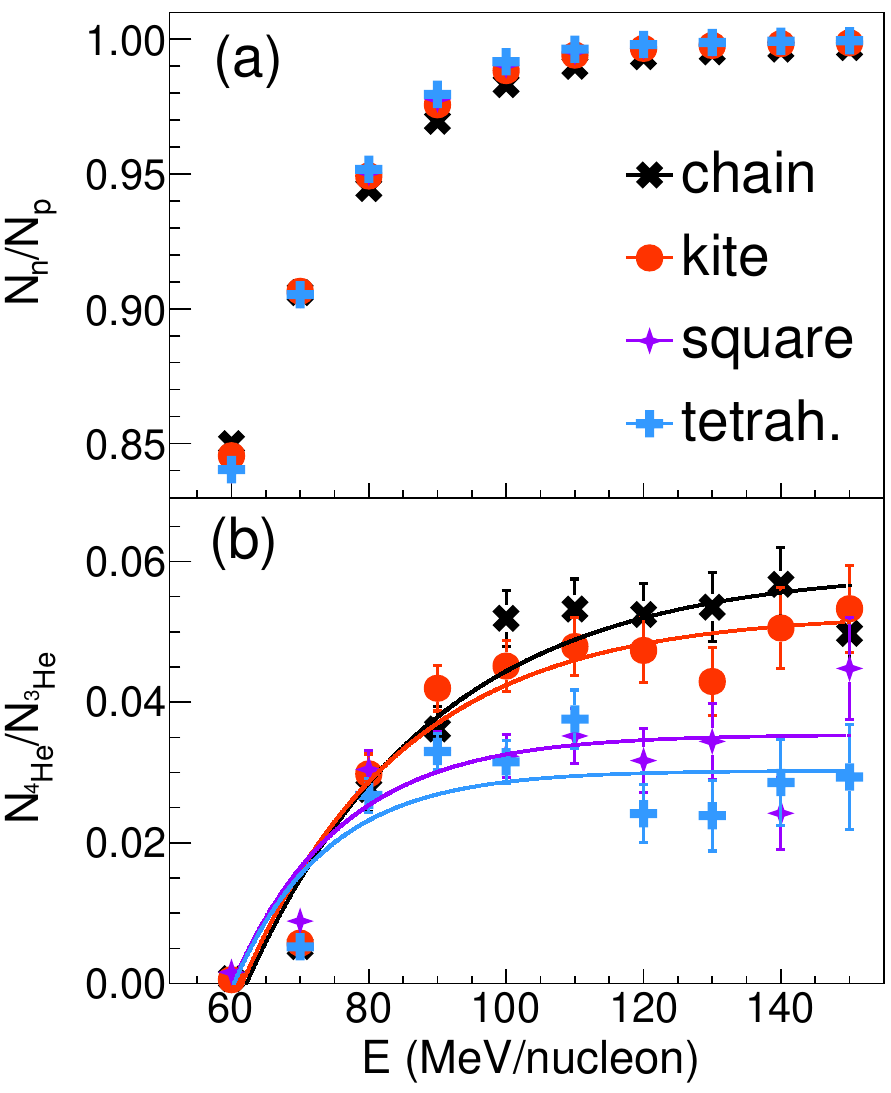}
\caption{Dependence of (a) N$_n$/N$_p$ and (b) N$_{^4He}$/N$_{^3He}$ on the incident energy when $^{16}$O has four different $\alpha$-cluster configurations.}
\label{fig8:R}
\end{figure}

\begin{figure}[htb]
\setlength{\abovecaptionskip}{0pt}
\setlength{\belowcaptionskip}{0pt}
\centering\includegraphics[scale=0.56]{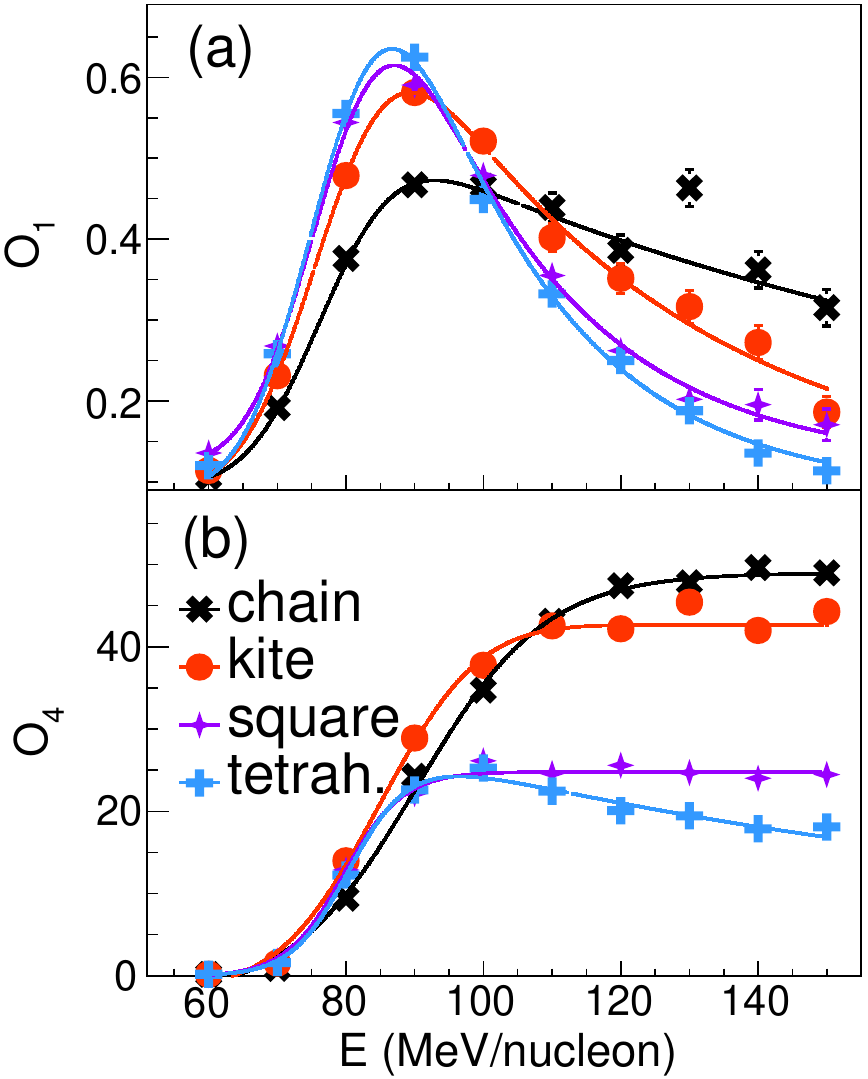}
\caption{Dependence of (a) O$_1$ and (b) O$_4$ on the incident energy when $^{16}$O has four different $\alpha$-cluster configurations.}
\label{fig7:O}
\end{figure}

In the EQMD model, the Pauli potential inhibits the system to collapse into the Pauli-blocked state at low energies and gives the model capability to describe $\alpha$-clustering. Before frictional cooling, the nucleon distribution of $^{16}$O is random, but after friction cooling it forms something like four-$\alpha$ configuration. For the four-$\alpha$ states of $^{16}$O, we have chosen four configurations: chain, square, kite and tetrahedron.  After the system goes long enough time till 
500 fm/c,  the final-state heavy fragments of which the excitation energy are greater than zero and the mass greater than 4 will be further deexcited by the GEMINI model. For a given $\alpha$-cluster configuration and incident energy point, the number of simulated events is 300,000. It should be noted that, for O$_1$ and O$_4$, the events when the denominator is zero are abandoned and only fill in the spectrum event by event with non-zero denominators.

\subsection{The effect of chain $\alpha$-clustering projectile with different polarization modes}
\label{Results01}

In this work, we refer to the plane formed by the intersection of the $x$ and  $z$ axes as the collision plane. Here, we polarize projectile with the chain of $^{16}$O both transversely and longitudinally, as shown in Fig.~\ref{fig1:transverse}. For other comparison case, the projectile is randomly rotated in four-$\pi$ solid angle. It can be imagined that the projection of the projectile on $x-y$ plane is only one $\alpha$-cluster point in the case of transverse polarization, while it is four $\alpha$-cluster points for the longitudinal polarization. In this way, different initial fluctuations among these three cases can be set and one can determine whether it has any effects on LGPT or not. Firstly, the yields of various types of fragments as a function of beam energy in  chain-like $^{16}$O bombarding on $^{40}$Ca collisions under three polarization modes are given, as shown in Fig.~\ref{fig3:yield_test}. One can see that the yields of proton and neutron increase with the increase of incident energy and reach stable values in energy region of 60 -- 150 MeV/nucleon. And the yields of deuteron, triton and $^3$He increase first and then decrease as incident energy increases.

For deuteron, triton and $^3$He, when the incident energy is less then 100 MeV/nucleon, their yields increase with the  incident energy, which is due to the fact that the composite system formed by $^{16}$O and $^{40}$Ca is in a state of fusion evaporation~\cite{NW03,NW05,NW06}.  At this stage, the compression and temperature of the collision system increase as incident energy increases. Thus it  evaporates more light clusters  etc.， such as proton, neutron, deuteron, triton and $^3$He~\cite{LWC03}.
However, with further increase of incident energy, the excitation energy of the system is so large that the system moves towards multiple fragmentation~\cite{NW03,NW05,NW06}. The phase-space volume occupied by proton and neutron becomes larger~\cite{LWC03}, which reduces the formation probability of deuteron, triton and $^3$He. These features have been observed in previous experiments~\cite{Nagamiya81}. In addition, for deuteron, triton and $^3$He, under the same conditions, the mass number is larger, the yield is smaller, which is consistent with the prediction from the thermal model~\cite{Bazak18}.

Different from the previous paragraph, for $^4$He, its yield starts at almost zero before 70 MeV/nucleon, then increases with the beam energy, and finally levels off or drops slightly (see  Fig.~\ref{fig6:yield}(f)). Moreover, the yield of $^4$He is about ten times that of $^3$He, which is exactly opposite to the prediction of the thermal model~\cite{Bazak18}.  The yield of $^4$He is greater than that of triton and $^3$He which can be attributed to the weaker Mott effect~\cite{Hagel12} on $^4$He than that on triton and $^3$He, i.e., a light nucleus would no longer be bound if the phase-space density of its surrounding nucleons is too large~\cite{Ropke82,Ropke83,WangR}.  This is because the $^4$He is well bound and compact while other light fragments is weakly bound and loose. Furthermore, from the trend of $^4$He yield, we speculate that $^4$He may be produced mainly through multiple fragmentation rather than fusion evaporation. At the beginning, when the incident energy is low, no multiple fragmentation has occurred, so the yield of $^4$He is almost zero. And with the increasing of incident energy, multiple fragmentation starts to occur and gradually dominates, so its yield increases. When the incident energy is large, it is difficult to decompose $^4$He due to the large binding energy, so its yield changes little or only slightly.

In Fig.~\ref{fig3:yield_test} (a) and (b), proton and neutron show insensitive to the polarization modes. However, for deuteron, triton, $^3$He, and $^4$He, they display obvious differences among longitudinal, transverse, and without polarization modes. And it is seen that deuteron shows more sensitive in low energy region, but it is opposite for triton, $^3$He, and $^4$He.

For the ratio of N$_n$/N$_p$ which is usually taken as a sensitive probe to neutron skin~\cite{XYS10,nskin,MaCW,WeiHL}, we can see from Fig.~\ref{fig5:R_test} (a) that it increases with the incident energy and eventually converges to 1, since the projectile and target are symmetric in this work. And there is no significant difference in the value of N$_n$/N$_p$ among different polarization modes. Additionally, as shown in Fig.~\ref{fig5:R_test}(b), the ratio of ${^4}$He to ${^3}$He has the similar trend with N$_n$/N$_p$ but has obvious difference for different polarization modes, and the curve is similar to the dependence of the yield of $^4$He on incident energy in Fig.~\ref{fig3:yield_test} (f), indicating that the change of the $^4$He yield is dominant.

Furthermore, ratios of O$_1$ and O$_4$ as a function of incident energy under different polarization modes (with different initial geometric fluctuations) are shown in Fig.~\ref{fig4:O_test} which could reflect nucleonic density fluctuation. One could expect that such geometric fluctuation has strong relation to the nucleonic density fluctuation. As mentioned above, the polarized projectile of chain-like $^{16}$O at longitudinal direction has larger geometric fluctuation than the transverse polarization one. And the geometric fluctuation for unpolarization one is between them. Here, one should notice that ratios of O$_1$ and O$_4$ are based on an equilibrium source. And the collision system at low energy could not reach equilibrium condition. Without such limit, one still can make the ratios by light nuclei but with less meanings. From Fig.~\ref{fig4:O_test} (a), one can see that the ratio of O$_1$ for unpolarization case has the largest value below 80 MeV/nucleon. As beam energy increases, however, the O$_1$ for longitudinal polarization gives the largest value and the one for transverse polarization shows the smallest which is as we expected. It shows that the initial-state geometric fluctuation of projectile with different $\alpha$-cluster configurations is sensitive to the O$_1$ at higher incident energies.  In Refs.~\cite{Liu2022,Deng2022}, density fluctuation is enhanced as beam energy or temperature increases which is associated with the LGPT in nuclear matter. In Fig.~\ref{fig4:O_test} (a), the ratios of O$_1$ can reach maximum value around 90 MeV/nucleon which depends on polarization modes. Such turning point could has physical meaning which may be associated with the LGPT and it will be cross-checked by charge distribution of the heaviest fragment below. For the ratios of O$_4$, it tends to be stable value as beam energy increases without turning value. But it seems that ratio of O$_4$ is sensitive to the polarization mode. Also one can see that trends of O$_1$ are similar to ones of the yield of triton, and trends of O$_4$ are similar to ones of the yield of $^4$He, from which we can infer that the yields of triton and $^4$He in the final-state product is more sensitive to geometric fluctuation. In addition, it can be seen from Fig.~\ref{fig3:yield_test},~\ref{fig5:R_test} and~\ref{fig4:O_test} that when the incident energy is low and the system is in the fusion evaporation stage, the yields and various ratios of different fragments are not sensitive to the geometric configuration of $^{16}$O, while they become sensitive only when the incident energy is high and the system is in the multiple fragmentation stage.

\subsection{The effect of projectile with different $\alpha$-clustering configurations}
\label{Results02}

Similarly in Sect.~\ref{Results01}, we first investigate the dependence of the yields of different types of fragments on incident energy with different $\alpha$-cluster configurations for $^{16}$O, the results of which are shown in Fig.~\ref{fig6:yield}. For proton and neutron, their yields increase with the incident energy. And they show no more difference among yields with different $\alpha$-cluster configurations. For deuteron, triton and $^3$He, their yields increase first and then decrease with the incident energy. And for $^4$He, its yield first increases and then becomes stable with the incident energy. Furthermore, when the incident energy is greater than 100 MeV/nucleon, the relationship among the yields of triton, $^3$He and $^4$He for $^{16}$O with different $\alpha$-cluster configurations is ``chain $>$ kite $>$ square $>$ tetrahedron'' and with an obvious difference.

\begin{figure}[htb]
\setlength{\abovecaptionskip}{0pt}
\setlength{\belowcaptionskip}{0pt}
\centering\includegraphics[scale=0.5]{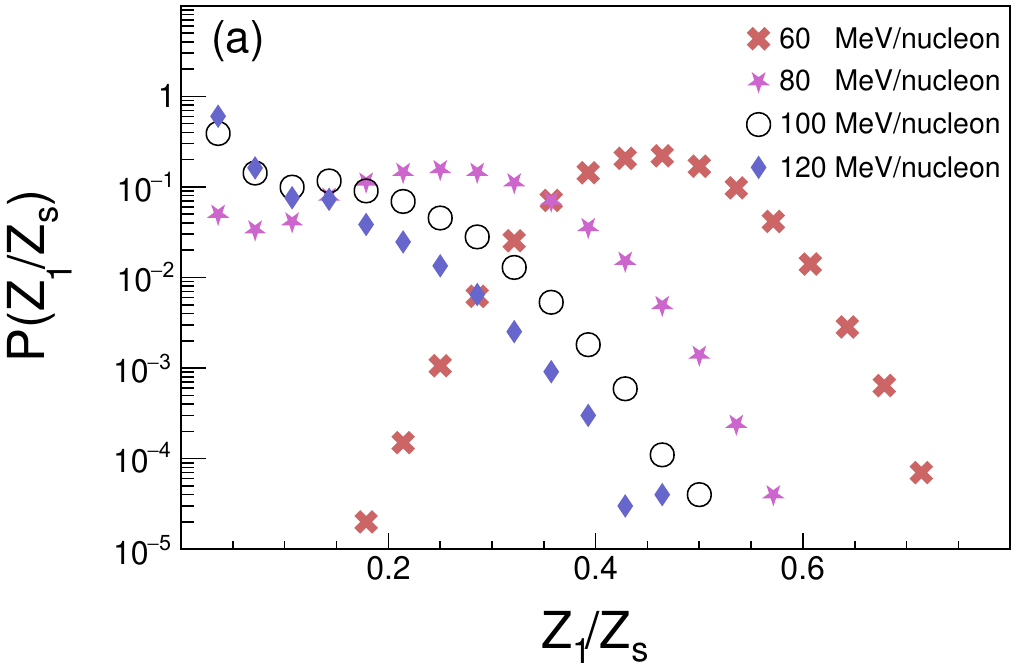}
\centering\includegraphics[scale=0.5]{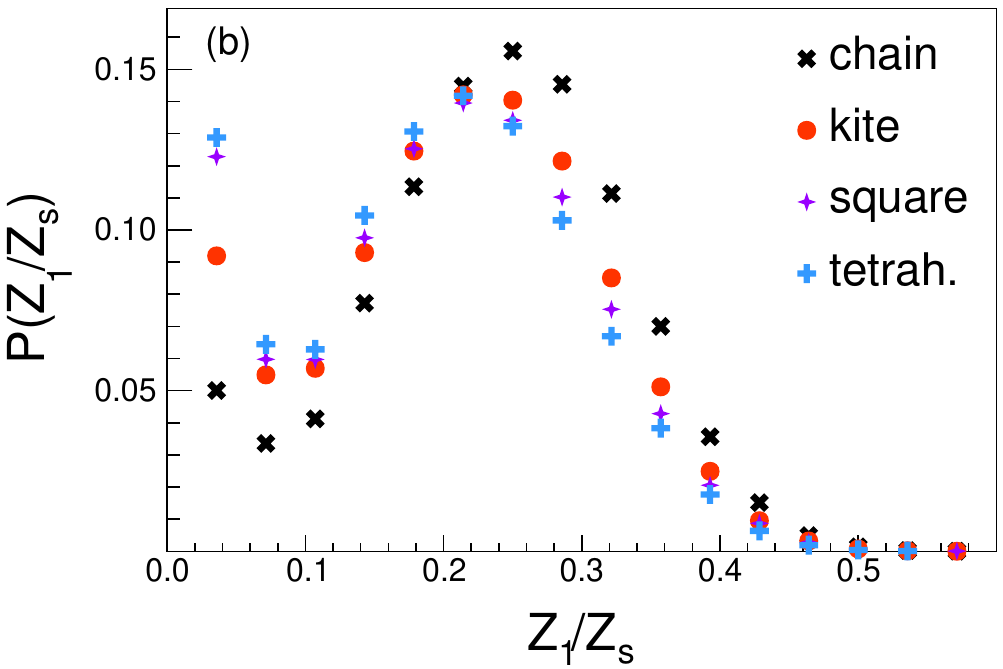}
\caption{Z$_1$/Z$_s$ distribution  for (a) chain-like $^{16}$O with different incident energies and (b) $^{16}$O with four different $\alpha$-cluster configurations when the incident energy is 80 MeV/nucleon.}
\label{fig9:bimodal}
\end{figure}

As shown in Fig.~\ref{fig8:R}, the trends of N$_n$/N$_p$ and N$_{^{4}He}$/N$_{^{3}He}$ are similar to those described in Sect.~\ref{Results01}. There is also no significant difference in the value of N$_n$/N$_p$ between different $\alpha$-cluster configurations as in Fig.~\ref{fig8:R} (a). The N$_{^4He}$/N$_{^3He}$ for chain-like configuration displays the largest values and the one for tetrahedron-like configuration is with the smallest value.

Ratios of O$_1$ and O$_4$ as a function of incident energy under different $\alpha$-cluster configurations are shown in Fig.~\ref{fig7:O}. For O$_1$, it first increases and then decreases with the incident energy. And below 100 MeV/nucleon, O$_1$ for chain-like configuration gives the smallest value and tetrahedron-like configuration is with the largest value. However, the hierarchy is opposite from 100 MeV/nucleon up to 150 MeV/nucleon. In addition, there are obvious peaks arising around 80 to 100 MeV/nucleon, which may be related to LGPT as mentioned above. For O$_4$, it first increases and tends to be stable with the incident energy except for the one with tetrahedron configuration slightly decreasing as beam energy increases after 100 MeV/nucleon. Moreover, the peak energy of O$_1$ is somehow different for various cluster configurations. And for O$_4$, the influence of different cluster configurations begins to appear at 80 MeV/nucleon and becomes stable after 100 MeV/nucleon.

As mentioned in Ref.~\cite{Lopez05}, the charge distribution of the heaviest fragment in intermediate energy heavy-ion collisions has been observed to be bimodal, which is expected as a generic signal of phase transition. So we plot the probability distribution for Z$_1$ over Z$_s$ for different incident energy and different $\alpha$-cluster configurations as shown in Fig.~\ref{fig9:bimodal}, where Z$_1$ is the charge of the heaviest fragment in each collision event and Z$_s$ is the sum of the charges of projectile and target. It can be clearly seen from Fig.~\ref{fig9:bimodal}(a) that for chain-like $^{16}$O, the probability distribution of Z$_1$/Z$_s$ starts to show a bimodal structure when the incident energy is greater than 80 MeV/nucleon, and this structure disappears until the incident energy is greater than 100 MeV/nucleon, further indicating that LGPT occurs within this incident energy range. Furthermore, as shown in Fig.~\ref{fig9:bimodal}(b), when the incident energy is 80 MeV/nucleon, the bimodal structure of the probability distribution curve corresponding to the square-like and tetrahedron-like projectile is the most obvious, followed by the kite-like, and the chain-like is the least obvious. Combined with the magnitude of geometric fluctuation for different $\alpha$-cluster configurations derived previously, it can be inferred that the larger the geometric fluctuation, the larger the incident energy resulting from LGPT, which can also be verified with the peak energy of O$_1$ in Fig.~\ref{fig7:O}(a).

\section{Conclusion}
\label{Conclusion}

The difference of geometric fluctuation caused by different $\alpha$-cluster configurations is mainly reflected in the effects on the yields of deuteron, triton, $^3$He and $^4$He, but it is dull for the yields of proton and neutron. By investigating the double ratios  $\mathcal{O}_\text{p-d-t}$ and $\mathcal{O}_\text{p-d-He}$ of light nuclei, we disclose that the magnitude hierarchy of geometric fluctuations is ``chain $>$ kite $>$ square $>$ tetrahedron'' for reactions of $^{40}$Ca induced by $^{16}$O with different $\alpha$-configuration. The maximum value of $\mathcal{O}_\text{p-d-t}$ is around 80 -- 100 MeV/nucleon which could be related to LGPT, and it is consistent with results from the charge distribution of the heaviest fragment in the same reaction. The current work sheds light on the effects of geometric fluctuation on LGPT in low-intermediate energy heavy-ion collisions.
In future, the yields of light nuclei produced in $^{40}$Ca + $^{16}$O central collisions with different incident energy can be measured through some experimental programs  in HIRFL at CSR, FRIB at MSU as well as other facilities. 
Since it was indicated in many previous studies that $^{16}$O in the ground state could be a tetrahedral 4$\alpha$ structure, we expect that the experimental data shall be compatible with the conclusions we have drawn in the previous sections for $^{16}$O with tetrahedral configuration.
Meanwhile, the yields of charged light nuclei are  intuitive and easily measurable physical quantities, and the single ratios of 
$^4$He/$^3$He as well as their double ratios $\mathcal{O}_\text{p-d-t}$ and $\mathcal{O}_\text{p-d-He}$ are better observables  since the insufficient detector's  effect in experiments can be cancelled, we expect the trend or saturation value of the excitation function of the ratios could give hints of geometric fluctuation.
Of course, collective observable, such as elliptic flow, may be also necessary for the further study on the phenomena discussed in this work.

\vspace{0.5cm}

Authors thank Dr. Kai-Jia Sun and Song Zhang for communications. This work was supported in part by the National Natural Science Foundation of China under contract Nos. 11890710, 11890714, 12147101, and 12205049, and the Guangdong Major Project of Basic and Applied Basic Research No. 2020B0301030008.

\end{CJK*}
\end{document}